\documentclass[aps,twocolumn,showpacs,preprintnumbers]{revtex4}

\usepackage[T2A]{fontenc}
\usepackage[utf8]{inputenc}
\usepackage[russian,english]{babel} 
\usepackage{graphicx}  
\usepackage{subfigure}
\usepackage{multirow}
\usepackage{longtable}
\usepackage{parskip}
\usepackage{dcolumn}   

\usepackage{bm}        
\usepackage{amsfonts}  
\usepackage{amsmath}   
\usepackage{amssymb}   
\usepackage{tikz}

\makeatletter
\let\original@widetext@top\widetext@top
\let\original@widetext@bot\widetext@bot
\def\widetext@top{0}
\def\widetext@bot{0}
\makeatother

\newcommand{\bra}[1]{\left\langle #1 \right|}

\newcommand{\pwisein}{\left\{ \begin{array}{ll}}
\newcommand{\pwiseout}{\end{array}\right.}
\newcommand{\ket}[1]{\left| #1 \right\rangle}

\begin{document}

\title{Influence of  Initial Entangled States on the Temperature-Dependent CHSH Inequality.}

\author{Esteban Marulanda}
 \email{esteban.marulandaa@udea.edu.co}
\author{Andrés Gómez}

\email{andres.gomez27@udea.edu.co}

\affiliation {\it Physics department, Universidad de Antioquia, Calle 70 No. 52-21, Medellín, Colombia}

\begin{abstract}  
We demonstrate that the temperature affects the validity of the CHSH inequality in an open bipartite two-qubit system. Specifically, for initial entangled states within the decoherence-free subspace (DFS), the CHSH inequality remains temperature-independent. In contrast, other entangled states exhibit a temperature threshold beyond which the inequality holds.
\end{abstract}

\date{\today}


\maketitle 

\section{Introduction}
\indent The advent of quantum information theory and the 50th-anniversary celebrations of the original publication of Bell's theorem in 2014 has only further elevated the importance of Bell inequalities in modern physics \cite{2022}. Bell inequalities have played a pivotal role in the foundational analysis of quantum mechanics, leading to insights that challenge our classical intuitions about the nature of reality. Introduced as part of Bell's Theorem in 1964, these inequalities stem from considerations of local causality, the idea that correlations between distant events should be explainable regarding local factors, such as the common source of the particles in question \cite{Bell1964}. Although initially confined to discussions among a small group of physicists and philosophers, interest in Bell's theorem has surged, particularly following the groundbreaking experiments by Aspect and colleagues. Nowadays, Bell inequalities are used to test entanglement in systems like photonic chips that generate entangled states \cite{Chen2021}. 
\newline
\indent However, those tests are primarily designed for systems with pure states. In reality, most systems interact with an environment, and generally, the state that describes the system of interest must be mixed. Therefore, the necessity for investigating quantum correlations and their dependence on environmental parameters is a task of vital importance. In the past years, much research has focused on this. To highlight a few,  \cite{Galve2010}, contrary to the conventional wisdom for a continuum variable setting, finds a high-temperature entangled system. \cite{corr_open} investigates various measures of quantum correlations in mixed states that interact with an environment through non-demolition and dissipative mechanisms. \cite{Marconi2022} presents an example of a system with a quantum correlation robust against noise. Finally, \cite{Fadel2018} addresses a system of N spins in thermal equilibrium, calculating a bound for the temperature of the multipartite Bell inequality.

While there is extensive literature on the topic, the relationship between the CHSH inequality for an open system and temperature remains unclear. Additionally, the possible underlying mechanisms that permit the existence of entangled states in high-temperature regimes. Could this phenomenon depend on the initial entangled state coupled to the environment?
This article aims to explore CHSH inequality for an open quantum system composed of two qubits and discuss the conditions for which an initial entangled state after coupling with an environment remains entangled even in the high-temperature regime. If an initial state doesn't maintain its entanglement in this high-temperature context, we then establish a specific temperature threshold beyond which the CHSH inequality is violated. 

This paper is organized as follows: Section 2 presents an example of a system defined by a Bell-type state. In the absence of external interactions, this system violates the CHSH inequality. We explore how interaction with a bosonic bath and temperature variation can affect such inequalities. Section 3 is dedicated to analyzing why some initial states have a violation of the CHSH inequality that is not affected by temperature. We conclude the study in Section 4.

\section{Methodology.}

Let us consider the version of the EPR paradox initially proposed by David Bohm \cite{Bohm1951-gh}. This involves a system of two spin-1/2 particles that move far apart enough that their mutual interaction becomes negligible. Among the possible final states, let us consider a Bell-type state,

\begin{equation}
\label{e1}
\ket{\psi_{AB}}=\frac{1}{\sqrt{2}}\left( \ket{0_A1_B}-\ket{1_A0_B}\right),
\end{equation}

\noindent where $A$ and $B$ represent the spin to be measured by Alice and Bob, respectively. Since both are unaware of the system's orientation in which the state \ref{e1} is prepared, they perform measurements of the received spin in an arbitrary basis rotated by $\theta$ and $\phi$ concerning the preparation reference frame. Considering a statistical ensemble of bell states given by \ref{e1},  the expected value for simultaneous measurement of A and B are \cite{Schumacher2010-vl}:


\begin{equation}
\label{e2}
\left \langle \hat{W}^{A}_\theta \hat{W}^{B}_\phi \right \rangle=-\cos{(\theta-\phi)},
\end{equation}

\noindent where \(\hat{W}^{j}_\theta= \sin{\theta}\hat{\sigma}_x^{j}+ \cos{\theta}\hat{\sigma}_z^{j}\), with \(j=A, B\) and \(\hat{\sigma}_x^{j}\), \(\hat{\sigma}_z^{j}\) represent the spin operators of $j$ in the directions indicated by the subscript. In the state specified by equation \ref{e1}, the CHSH inequality is violated for certain specific values of \(\theta\) and \(\phi\),

\begin{equation}
\label{e3}
\begin{split}
S & =\left \langle \hat{W}^{A}_0 \hat{W}^{B}_{\pi/4} \right \rangle-\left \langle \hat{W}^{A}_0 \hat{W}^{B}_{3\pi/4} \right \rangle+ 
\left \langle \hat{W}^{A}_{\pi/2} \hat{W}^{B}_{\pi/4} \right \rangle \\ &+ \left \langle \hat{W}^{A}_{\pi/2}  \hat{W}^{B}_{3\pi/4} \right \rangle=-2\sqrt{2}<-2.
\end{split}
\end{equation}

This example highlights a well-established understanding: a theory based on Bell locality cannot account for the results of quantum mechanics in every instance \cite{sep-bell-theorem}. Nonetheless, in the context of open quantum systems, we often examine the reduced system due to the impracticality of achieving complete isolation. Consequently, questions arise regarding the sufficient conditions for a reduced system to violate the CHSH inequality.

\begin{figure}[h]
\centering
\begin{tikzpicture}
    \filldraw (0,0) circle (0.2cm);
    \filldraw (2,0) circle (0.2cm);

    \draw[->] (0.7,1.6) -- (0.3,0.5);
    \draw[->] (1.3,1.6) -- (1.7,0.5);
    \node at (1,1.85) {Entangled};
    
    \draw[->] (-0.7,0) -- (-1.3,0) node[left] {A};
    \draw[->] (2.7,0) -- (3.3,0) node[right] {B};

    \draw[samples=250,domain=0:2, smooth] plot (\x,{0.1*exp(-10*(\x-1)*(\x-1))*sin(50*\x r)});

    \draw[samples=150,domain=0:-2, variable=\y, shift={(0,0)}] plot ({0.1*sin(10*\y r )+0.25*\y},\y);
    \draw[samples=150,domain=0:-2, variable=\y, shift={(2,0)}] plot ({-0.1*sin(10*\y r)-0.25*\y},\y);
    \draw (2.6,-2.25) circle (0.25cm);
    \node at (2.6,-2.25) {T};
    \draw (-0.6,-2.25) circle (0.25cm);
    \node at (-0.6,-2.25) {T};
    
    \node at (-0.8,-1) {\( g_k \)};
    \node at (2.8,-1) {\( g_k \)};

\end{tikzpicture}
\label{compz}
\caption{Schematic showing the entangled state of two distant spins (measured by Alice and Bob) with negligible interaction, coupled to a thermal bath at temperature $T$ via $g_k$.}
\end{figure}
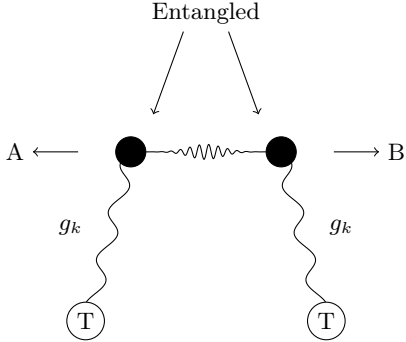

To observe this, let us consider a simplified experiment illustrated in Figure 1: two spins that have interacted become entangled at a moment \( t<0 \),  at \( t=0 \) they couple to a thermal bath consisting of bosons such that the interaction is now solely between the spins and the bath. In this scenario, we have the following initial state of the entire system, composed of the pair of spins and the bath:

\begin{equation}
\label{e4}
\hat{\rho}(0)=\hat{\rho}_S(0)\otimes \hat{\rho}_E,
\end{equation}

\noindent in this expression, \(\hat{\rho}_S(0)=\ket{\psi} \bra{\psi}\) represents the density matrix of the bipartite system composed of two spins whose Hamiltonian is given by \(\hat{H}_S=\sum_{m\in\{A,B\}}\frac{\omega^{m}_0}{2}\hat{\sigma}^{m}_z\), where \(\omega^{m}_0\) is the energy difference between the eigenstates \(\ket{0_m}\) and \(\ket{1_m}\). \(\hat{\sigma}^{m}_z\) is the spin operator in the z-direction for spin \(m\). On the other hand, \(\hat{\rho}_E=\frac{\exp(-\beta \hat{H}_E)}{Z_E}\) characterizes the thermal state of the bosonic bath, with a Hamiltonian defined by \(\hat{H}_E=\sum_s \omega_s \hat{b}^{\dagger}_{s}\hat{b}^{}_{s}\). The interaction is assumed to be of the oscillator type, that is,

\begin{equation}
\label{e5}
\hat{H}_{int}=\frac{1}{2}\sum_{m\in\{A,B\}}\sum_s \hat{\sigma}^{m}_z\left( g_s \hat{b}^{\dagger}_{s}+g^{*}_s \hat{b}_{s} \right),
\end{equation}

\noindent where \(g_s\) represents the coupling constants. Physically, this model represents a situation of decoherence without dissipation, as \(\left [ \hat{H},\hat{\sigma}^{m}_z \right ]=0\). Our focus now shifts to calculating equation \ref{e2} for the bipartite system, that is,


\begin{equation}
\label{e7}
\begin{split}
 & \left \langle \hat{W}^{A}_\theta \hat{W}^{B}_\phi \right \rangle_S=\text{tr}_S\left( \hat{\rho}_S(t) \hat{W}^{A}_\theta \hat{W}^{B}_\phi \right)= \\ &
=\sin{\theta}\sin{\phi} \left( \rho_S^{1100}(t)+\rho_S^{0011}(t)+
\rho_S^{0101}(t)+\rho_S^{1010}(t)
 \right) \\&
+\sin{\theta}\cos{\phi} \left( \rho_S^{0100}(t)+\rho_S^{0010}(t)-
\rho_S^{1101}(t)-\rho_S^{0111}(t) \right) \\&
+\cos{\theta}\sin{\phi} \left( \rho_S^{1000}(t)+\rho_S^{0001}(t)-
\rho_S^{1101}(t)-\rho_S^{0111}(t) \right) \\&
+\cos{\theta}\cos{\phi} \left( 
\rho_S^{0000}(t)+\rho_S^{1111}(t)-
\rho_S^{1001}(t)-\rho_S^{0110}(t) \right),
\end{split}
\end{equation}

\noindent here, \(\hat{\rho}_{S}(t)=\text{tr}_E\left(\hat{U}(t) \hat{\rho}(0) \hat{U}^{\dagger}(t) \right)\) represents the reduced density matrix of the bipartite system and $\rho_S^{ijkl}(t)=\bra{ij}\hat{\rho}_S(t)\ket{kl}$. The computation of the matrix elements has been studied in the literature when the coupling with a single spin is considered \cite{heinz}. However, since there is no interaction between the spins, it is feasible to generalize the result for two spins, thus obtaining,

\begin{equation}
\label{e8}
\begin{split}
\bra{ji}\hat{\rho}_S(t) \ket{lk} & = \rho_S^{jilk}(0) \times \\ & \text{tr}_E[ \exp{\left(  \frac{1}{2}p_{ij} \sum_s  \left( \alpha_s\hat{b}^{\dagger}_{s}-\alpha^{*}_s \hat{b}_{s} \right)\right)}\hat{\rho}_E(0) \\ & \exp{\left(  \frac{1}{2}p_{lk} \sum_s  \left( \alpha^{*}_s \hat{b}_{s} -\alpha_s\hat{b}^{\dagger}_{s}\right)\right)}] ,
\end{split}
\end{equation}

\noindent we have defined \(\alpha_k=2g_k\frac{1-\exp{(i\omega_kt)}}{\omega_k}\) and also \(p_{ij}=(-1)^{i}+(-1)^{j}\). Thus, for the state $\ket{\psi}=\ket{\psi_{AB}}$ given by equation \ref{e1}, only the matrix elements $\rho_S^{jilk}(0)$ that satisfy \(l \neq k\) and \(i \neq j\) are non-zero. Therefore, we observe no temperature dependence in the bipartite system at any point for this initial state since equation \ref{e7} matches equation \ref{e2}; hence, the Bell inequalities will be violated at any temperature. 

Now, among the possible initially entangled states, consider \(\ket{\psi_{AB}}=\frac{1}{\sqrt{2}}\left( \ket{1_A1_B}+\ket{0_A0_B}\right)\), which violates the CHSH inequality when considering the same \(\theta\) and \(\phi\) from equation \ref{e3}. For the new bell-type state, we obtain \textbf{for} the reduced system under the same physical conditions as in the previous case,
 

\begin{equation}
\label{e8}
\begin{split}
\left \langle \hat{W}^{A}_\theta \hat{W}^{B}_\phi \right \rangle_S  & = \cos{(\theta-\phi)}\times \\ & \times \exp{\left(-8\sum_s  \left | \alpha_s \right |^{2} \text{coth}\left(\frac{\omega_s \beta}{2}\right)\right)}.
\end{split}
\end{equation}

Moreover, in this case, equation \ref{e3} exhibits a temperature dependence for the bipartite system. Explicitly,

\begin{equation}
\label{e9}
\begin{split}
S=2\sqrt{2} \times \exp{\left(-8\sum_s  \left | \alpha_s \right |^{2} \text{coth}\left(\frac{\omega_s \beta}{2}\right)\right)}.
\end{split}
\end{equation}

This expression demonstrates how the Bell inequality would be satisfied as a function of temperature and the parameters associated with the thermal bath and interaction. For instance, if we assume that the number of bosons tends to infinity, equation \ref{e9} can be expressed in terms of the spectral density of the thermal bath as,

\begin{equation}
\label{e10}
\begin{split}
S &=2\sqrt{2} \times  \\ & \times \exp{\left(-32\int_{0}^{\infty} d \omega \hspace{0.1cm}J(\omega)\left | \frac{\sin{\left(\frac{\omega t}{2}\right)}}{\omega/2}\right |^{2} \text{coth}\left(\frac{\omega \beta}{2}\right)\right)},
\end{split}
\end{equation}

\noindent where, \(J(\omega)=\sum_s \left | g_s \right |^{2} \delta(\omega_s-\omega)\). For this example, we see that given the spectral density of the bath, we can find an upper limit for the temperature \(T_c\) such that \(S<2\). Let's consider the thermal limit \(t\rightarrow \infty\), in this case, \(\sin^{2}{\left(\frac{\omega t}{2}\right)}/\omega^{2}\rightarrow \pi t \delta(\omega)\) \cite{Bransden1982-mv}. Commonly spectral densities satisfy \(J(\omega=0)=0\) \cite{2007a}. Then, we have,

\begin{equation}
\label{e11}
\begin{split}
S & =2\sqrt{2} \times \exp{\left(-256  k_b T\frac{\mathrm{d}J}{\mathrm{d} \omega} | _{\omega=0} \right)} \\ & \Rightarrow T_c=\frac{\text{ln}\sqrt{2}}{256 k_b  \frac{\mathrm{d}J}{\mathrm{d} \omega} | _{\omega=0}}.
\end{split}
\end{equation}

The upper equation shows the expected physical situation. In the high-temperature limit, the CHSH inequality is anticipated to hold since \(S\rightarrow 0\). On the other hand, the equation below indicates the critical temperature dependence on the spectral density.


\section{Discusion.}

In the previous section, we prove that CHSH inequality could be independent on the temperature once the system interacts with a bosonic bath for certain initial entangled states of the bipartite system. We find out that this is because the system's initial states belong to the decoherence-free subspace (DFS), which is not influenced by the bath \cite{Lidar1999}. That is, the density matrix associated with the bipartite system preserves unitary evolution $\hat{\rho}_S(t)=\hat{U}_S  \hat{\rho}_S(0) \hat{U}_S^{\dagger}$ with $\hat{U}_S=\text{exp}(-it\hat{H}_S)$. In our previous example, states belonging to DFS are given by those states which are eigenstates of $\frac{1}{2}\sum_{m \in \{A, B\}}\sum_s \hat{\sigma}^{m}_z$. In short, initial entangled states satisfying this condition as in the case of equation \ref{e1} violated CHSH inequality independent of temperature. However,  other states, such as in the case \(\ket{\psi_{AB}}=\frac{1}{\sqrt{2}}\left( \ket{1_A1_B}+\ket{0_A0_B}\right)\), will exhibit a critical temperature beyond which the Bell inequality is satisfied. Therefore, the environment constrains the validity of using a Bell test as a measure of entanglement. That is, violating the Bell inequality is a sufficient condition to ensure entanglement but not a necessary one \cite{Werner1989}. In this scenario, employing other more effective entanglement tests is essential.

\section{Conclusion.}

 We show that starting from an initial state that violates the CHSH inequality, we ascertain that states residing in the Decoherence-Free Subspace (DFS) allow two qubits, when interacting with an oscillator environment, to consistently violate the CHSH inequality irrespective of temperature. For states outside the DFS, we determine a temperature threshold beyond which the CHSH inequality is no longer violated. This threshold holds true across various spectral densities of the environment. 

\newpage
\begin{widetext}
\bibliographystyle{unsrt}
\bibliography{refs}
\end{widetext}

\end{document}